\documentstyle[12pt,epsbox]{article}

\begin{document}

\pagenumbering{arabic}

\begin{center}

{\bf Concept of new gamma ray detector}

{Satoko Osone}

{\it Matsudo 1083-6 A-202, Matsudo, Chiba 271-0092, Japan}

{osone@icrr.u-tokyo.ac.jp}
\end{center}

\section*{Abstract}

We present a concept of a new gamma ray detector in order to observe undetected TeV gamma ray background. 
We measure a track of an electron-positron pair made by a pair creation in a magnet.
By using Si as a tracker in a magnetic field 3 T, an energy range is up to 10 TeV.

\section{Introduction}
Gamma ray from point sources and diffuse gamma ray background below 10 GeV have been observed on the satellite and that below 100 GeV is scheduled to observe with {\it GLAST}.
Gamma ray from point sources above several 100 GeV have been observed on a ground array as a cherenkov light.
Diffuse gamma ray background above 100 GeV have not been observed. 
Because we can not distinguish between a diffuse gamma ray induced air shower and a proton induced air shower on a ground. Only upper limit is obtained(eg. Aharonian et al. 2002 ). Also because statistics of point sources above 100 GeV on a satellite is low  and there is a limit on a weight on a satellite.
Statistics of TeV diffuse gamma ray background is up to 10 TeV on a satellite.
Therefore, we present a concept of a new light detector on a satellite  to observe TeV diffuse gamma ray emission up to 10 TeV.

\section{Physics of TeV gamma ray diffuse background}
Origin of  diffuse galactic gamma ray background is considered as a
gamma ray product of an interaction between an accelerated electron and an interstellar medium.
Galactic diffuse gamma ray background in the energy range from 100 MeV to 10 GeV is well explained with a sum of
a bremsstrahlung and an inverse compton by an accelerated electron  and a nucleon nucleon production and isotropic diffuse emission(Hunter et al., 1997).
At TeV energy range, an inverse compton process by an accelerated electron is dominated(Porter \& Protheroe 1997, Tateyama \& Nishimura 2001) as figure 1.
We can obtain an index of an electron injection energy spectra and a maximum energy of an electron from an energy spectra of diffuse galactic TeV gamma ray background.

Origin of  diffuse extragalactic gamma ray background is considered as  unresolved point sources. The candidate is Blazar.
The total flux of extragalactic diffuse gamma ray background up to 10 GeV is well explained by unresolved Blazars(Stecker \& Salamon 1996).
We can obtain a degree of a fluctuation of  diffuse extragalactic gamma ray background's flux of a field to a field.
This value is $\sqrt{N}/N$ for a number of unresolved sources $N$.
We can obtain indirectly a number of unresolved point sources $N$ below a sensitivity of a detector at TeV gamma ray range.
This method was used for a non-imaging detector as Xray satellite {\it GINGA}.
This give a perspective for a TeV gamma ray experiment on a ground which search for a point sources.

The events beyond Greisen-Zatsepin-Kuz'min cut-off(Greisen 1966, Zatsepin \& Kuz'min 1966) around $5\times 10^{19}$eV in the energy spectrum of cosmic ray have been found with AGASA(Takeda et al. 1998, Sakaki et al. 2001).
This result give an opportunity for a Top down scenario.
There is gamma ray as a decay product from a massive X particle( Protheroe \& Stanev 1996, Sigl 1996) as figure 2.
From a flux of diffuse extragalactic gamma ray background with {\it EGRET}, a mass of X particle is limited to lower than $10^{13}$ eV. 
By giving a flux of diffuse extragalactic gamma ray background up to 10 TeV, there is a new limit to a physical quantity of X particle.

\begin{figure}[htb]
 \begin{center}
\psbox[height=7cm]{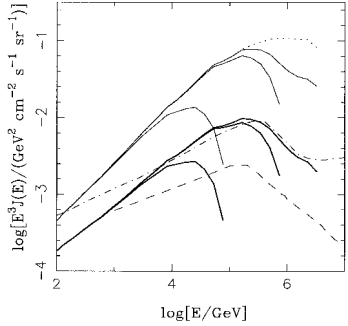}
 \end{center}
\caption{The expected diffuse galactic gamma ray background(Porter \& Protheroe 1997).
Thick solid curve show an inverse compton spectrum for $E^{-2.4}$ electron spectrum, thin solid curve show the inverse compton spectrum for $E^{-2.2}$ electron spectrum.
The lowest curve is for a cut-off at 100 TeV, next higher branch is for a cut-off at 1 PeV, the next higher brand is for no cut-off.
The each curve include an attenuation on the cosmic microwave background.}
\end{figure}

\begin{figure}[htb]
 \begin{center}
\psbox[height=7cm]{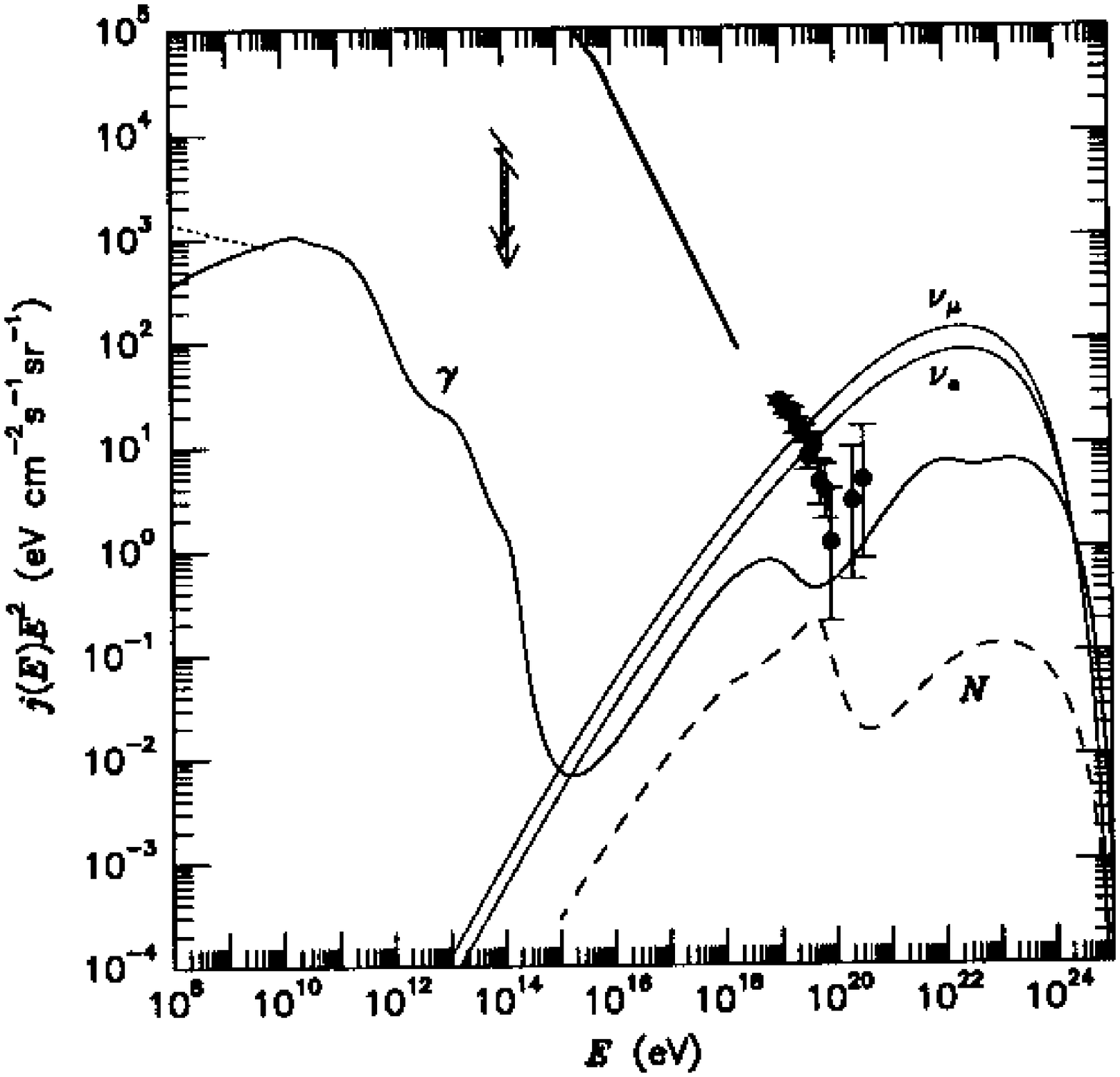}
 \end{center}
\caption{An energy spectrum of a decay product of X particle(Sigl 1996). The dot indicates  diffuse extragalactic gamma ray background with {\it EGRET}.}
\end{figure}

\section{Concept of detector}

Original technique to observe gamma ray on satellite is
to measure an energy of gamma ray by absorbing all energy with an absorber and measure an arrival direction with a tracker as {\it EGRET}, {\it GLAST}.
To obtain a large stopping power for TeV gamma ray, we have to prepare a weight absorber.
Hence, we have a idea which need no weight absorber in order to observe TeV gamma ray.
By occurring a pair creation with Pb, we convert gamma ray into an electron-positron pair.
And, we measure a track of an electron-positron in a magnet and obtain an energy of an electron-positron with a relation $p(GeV/c)\sim 0.3ZB(T)R(m)$. Here, B is a magnetic field, R is a curvature radius. And we obtain an arrival direction of a gamma ray from two tracks of an electron-positron.
Also, we can discriminate between a gamma ray induced pair and a cosmic ray by a number of a track.

Actual use of a magnet in the sky have been done as BESS, AMS in order to observe anti particles. The weight of a magnet is 600 kg and the weight of Pb is 97 kg for surface area 1 $m^2$ and a thick of 1 radiation length(5.5 mm).
Therefore, This detector is lighter than an original detector for a same energy range.

We show an overview of detector at figure 3.
The detector is $\sim$ 1m diameter and has a cylindrical structure.
We use Si of a pitch length 10$\mu$m as a tracker.
Then, we have a sensitivity up to 10 TeV and arrival resolution 10$^{-5}$ rad.
With an arrangement of Si as cylinder, we obtain a large field of a view about 90$^o\times 180^o$ and can use for all sky monitor below 100 GeV.

We have 3 sigma photons (9 photons) at 100 GeV for a point source which has a flux of 1 crab with an effective area 1 m$^2$ and 1 year observation.
Therefore, 100 GeV is a maximum energy for an observation of a point source.
As for point sources, we have a same sensitivity with {\it GLAST} though 
our detector have a better angular resolution because statistics of point sources is low.

We calculate statistics of diffuse gamma ray background with this detector.
By assuming a photon index 2.0 with no cut-off up to 10 TeV against a flux detected with {\it EGRET}(Hunter S.D. et al. 1997, Sreekumar et al. 1998), we expect 40 photons, 100 photons at 10 TeV for  diffuse extragalactic gamma ray background, diffuse galactic gamma ray background respectively, with 1 year observation, an effective area 1$m^2$ and a field of view $90^o\times180^o$.
We expect 30 photons, 2 photons at 10 TeV for  diffuse galactic gamma ray background from an electron spectrum of an injection index 2.2, 2.4, respectively with 1 year observation, an effective area 1$m^2$ and a field of view $90^o\times180^o$, from figure 1.

We check if we obtain a clear track in a magnet.
The deflection angle in a magnet is
\begin{equation}
\phi_{mag} = \frac{LeB}{pc}=\frac{300 B(T)L(m)}{pc(MeV)}
\end{equation}
Here, $L$ is a traverse length, $B$ is a magnetic field, $pc$ is a momentum of a particle.
The deflection angle by a multiple scattering is 
\begin{equation}
\phi_{scatt}= \frac{E_s}{\sqrt{2}p\beta c}\sqrt{\frac{l}{X_0}}
\end{equation}
Here, $E_s$=21 MeV, $X_0$ is radiation length, $l$ is the thick of a material.
The ratio of these two deflection angle is,
\begin{equation}
\frac{\phi_{scatt}}{\phi_{mag}}= \frac{0.05}{B(T) \beta L(m)}\sqrt{\frac{l(m)}{X_0(m)}}
\end{equation}
For a relativistic particle ($\beta=1$), $B=3 T, L=1 m, X_0$=0.09 m(Si), $l=N\times500\mu m=2\times10^{-3}$ m for a number of an interaction point $N$=4, $\phi_{scatt}/\phi_{mag}$=0.002.
Therefore, we can detect a clear track of electron-positron pairs.

Next, we confirm a resolution of a momentum.
The resolution of a momentum  from a track measurement error(Kleinknecht 1986) is
\begin{equation}
\frac{\Delta P}{P}= \frac{\sigma(x)(m)p(GeV/c)}{0.3B(T)L(m)^2}\sqrt{\frac{720}{(N+4)}}
\end{equation}
Here, $\sigma(x)$ is a precision of a position.
We obtain 10 \% for $B=3 T, \sigma(x)=10^{-5} m, L=1 m, p=1 TeV/c, N=4$.

\begin{table}
\begin{center}
\caption{List of a performance}
\begin{tabular}{cc}\hline
energy range for Background    & $\sim$10 TeV \\
             for point source &  $\sim$100 GeV \\
angular resolution & $10^{-5}$ rad \\
F.O.V & 90$^o\times180^o$ \\
effective area & 1$m^2$ \\
momentum resolution at 1 TeV & 10\% \\ 
magnetic filed & 3 T \\ \hline
\end{tabular}
\end{center}
\end{table}

\begin{figure}[htb]
 \begin{center}
\psbox[height=15cm]{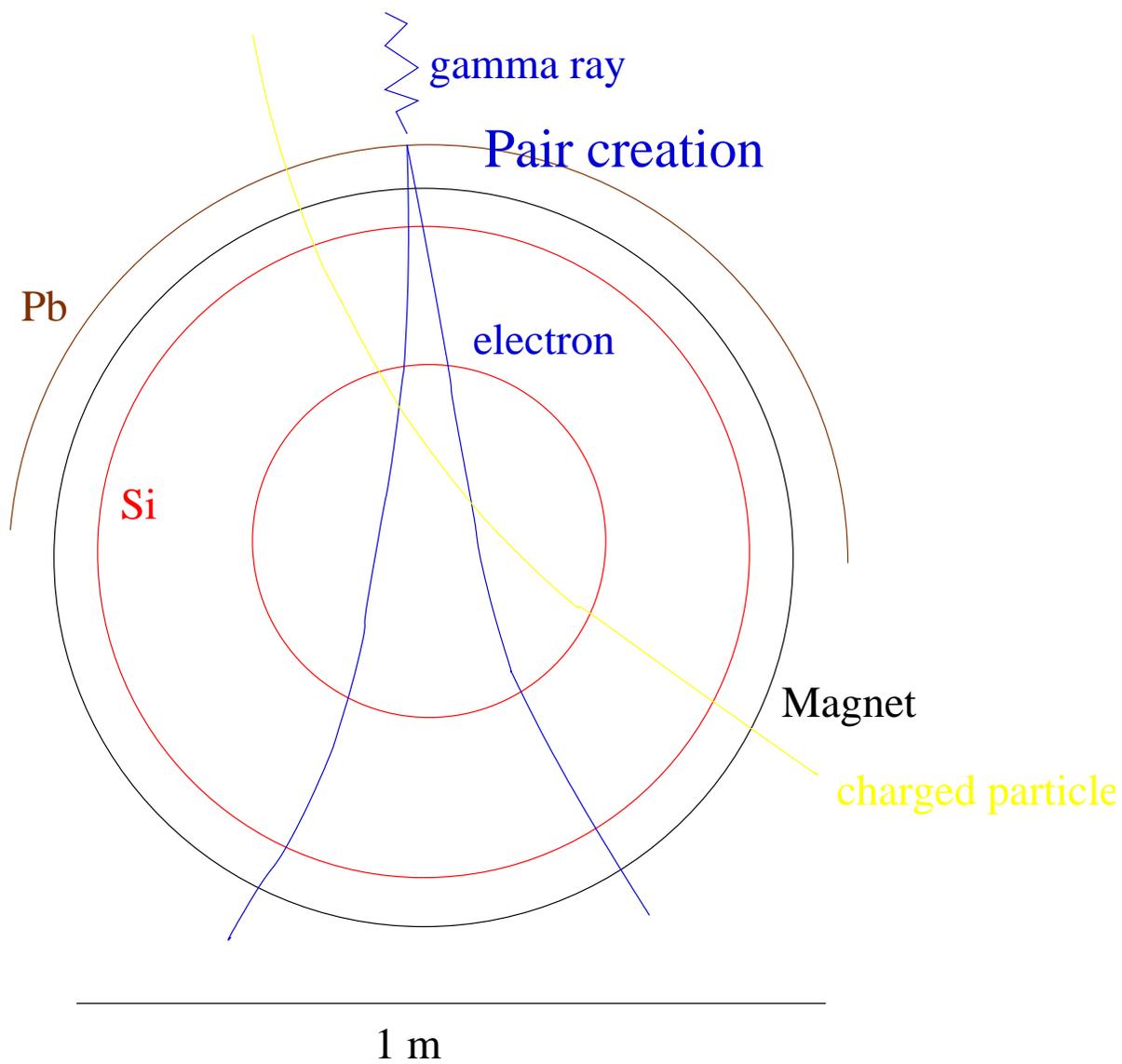}
 \end{center}
 \caption{Overview of new gamma ray detector}
\end{figure}

\section{References}

\begin{itemize}
\setlength{\itemsep}{-1.5mm}
\setlength{\itemindent}{-8mm}
\item[]1. Aharonian F.A. et al., 2002, Astropart. 17, 459.
\item[]2.Greisen K., 1966, Phys.Rev.Lett., 16, 748.
\item[]3.\ Hunter S.D., 1997, ApJ, 481, 205.
\item[]4. Kleinknecht K., 1986, '{\it Detectors for particle radiation}',  Cambridge University Press.
\item[]5. Perkins D.H., 2000, '{\it Introduction to High Energy Physics}', 4th edition, Cambridge University Press.
\item[]6. Porter T.A. \& Protheroe R.J., 1997, J.Phys.G, 23, 1765.
\item[]7.\ Protheroe R.J.\&Stanev T., 1996, Phys.Rev.Let., 77, 18, 3708.
\item[]8. Sakaki N. et al., 2001,Proc. 27th Int. Cosmic Ray Conf.(Hamburg),HE 1, 333.
\item[]9.\ Sigl G., astro-ph/9611190.
\item[]10.\ Sreekumar P., 1998, ApJ, 494, 523.
\item[]11. Stecker F.W.\&Slamon M.H., ApJ, 1996, 464, 600.
\item[]12. Takeda M. et al., 1998, Phys. Rev. Lett, 81, 6, 1163.
\item[]13. Tateyama N.\& Nishimura J., 2001, Proc. 27th Int. Cosmic Ray Conf.(Hamburg),6, 2343.
\item[]14.Zatsepin G.T.\& Kuz'min V.A., 1966, Zh. Eksp. Teor.Fiz., 4, 114.
\end{itemize}

\end{document}